\documentclass[12pt]{iopart}

\usepackage{amsfonts,amssymb,graphicx,multicol}

\newcommand{\Eq}[1]{Eq.~(\ref{#1})}

\newcommand{\beq}{\begin{equation}}
\newcommand{\eeq}{\end{equation}}
\newcommand{\ua}{\uparrow}
\newcommand{\da}{\downarrow}

\begin{document}

\title[Interaction quench dynamics in the Kondo model with local magnetic field]{Interaction quench dynamics in the Kondo model in presence of a local magnetic field}

\author{M. Heyl and S. Kehrein}

\address{Department of Physics, Arnold Sommerfeld Center for Theoretical Physics and Center for NanoScience, Ludwig Maximilians Universit\"at M\"unchen, Theresienstr. 37, 80333 Munich, Germany}

\ead{Markus.Heyl@physik.lmu.de}

\begin{abstract}

In this work we investigate the quench dynamics in the Kondo model on the Toulouse line in presence of a local magnetic field. It is shown that this setup can be realized by either applying the local magnetic field directly or by preparing the system in a macroscopically spin-polarized initial state. In the latter case, the magnetic field results from a subtlety in applying the  bosonization technique where terms that are usually referred to as finite-size corrections become important in the present non-equilibrium setting. The transient dynamics is studied by analyzing exact analytical results for the local spin dynamics. The time scale for the relaxation of the local dynamical quantities turns out to be exclusively determined by the Kondo scale. In the transient regime, one observes damped oscillations in the local correlation functions with a frequency set by the magnetic field.

\end{abstract}

\pacs{72.15.Qm,71.27.+a}
\submitto{\JPC}
\maketitle

\section{Introduction}

In recent years the possibility to experimentally study non-equilibrium dynamics in quantum many-body systems stimulated theoretical investigations of various model systems out of equilibrium. Of particular interest for this work is the observation that quantum dots can act as magnetic impurities displaying Kondo physics.\cite{Exp_Kondo} Due to the flexibility to vary system parameters in time by applying unscreened electrical or magnetic fields, quantum dots offer the framework for the experimental investigation of non-equilibrium dynamics in quantum impurity systems.

The Kondo model describes the physics of a localized spin $1/2$ coupled to a fermionic bath via an exchange interaction. At low temperatures, the sea of conduction band electrons develops a spin polarization cloud, the so-called Kondo cloud, providing a mechanism to screen the local magnetic moment. In the zero temperature limit, the screening becomes dominant leading to the emergence of the Kondo singlet such that the spin-polarization cloud is bound to the local spin with an associated binding energy $T_K$, the Kondo temperature. The Kondo effect manifests itself most prominently in the Kondo resonance, a sharp peak in the local density of states that is pinned at the Fermi energy. 

As the Kondo model is the paradigm model for strongly correlated impurity systems, it is of particular interest that for a certain line in parameter space, the Toulouse limit, the Kondo model becomes exactly solvable by a mapping onto a quadratic resonant level model.~\cite{Toulouse,Leggett} In equilibrium as well as for an interaction quench, the Toulouse limit describes correctly many generic and universal properties of the Kondo model~\cite{Leggett,Lob_dyn,Lob_eff} such as the local spin dynamics that is also investigated in this work. Due to the complexity of time-evolution for systems out of equilibrium it is instructive to investigate those particular examples where exact and nonperturbative solutions are accessible.

Most of the work on time-dependent non-equilibrium systems has been concentrating on interaction quenches.~\cite{Lob_dyn,Lob_eff,Nordl_Kondo_develop,Anders,Kondo_neq,Mitra,Lesage,quenches,Komnik} Due to its paradigmatic importance, the non-equilibrium quench dynamics in the Kondo model or the related Anderson impurity model in the local moment regime has been analyzed in a number of works.~\cite{Lob_dyn,Lob_eff,Nordl_Kondo_develop,Anders,Kondo_neq,Mitra,Lesage} Here, the transient dynamics in the Kondo model will be investigated for an interaction quench in presence of a local magnetic field. It will be shown that this scenario may be realized by either applying a local magnetic field directly or by preparing the system in a state in which the conduction band carries a macroscopic spin polarization. As the Kondo effect is sensitive to spin degeneracy, the local magnetic field is expected to influence the properties in the Kondo model considerably. The analysis of the transient dynamics and the emergence of the steady state will focus on the local spin dynamics as it is directly affected by the local magnetic field.

In the non-equilibrium setting where the lead carries a macroscopic initial spin polarization a subtlety in the application of the bosonization technique arises. Under certain circumstances as in this setting, terms that are usually referred to as finite-size corrections may turn out to be relevant in the non-equilibrium case. As a consequence of the ``finite size contributions'' the spin-polarized initial state effectively acts as a local magnetic field applied to the local spin.

This paper is organized as follows. In Sec.~II, the effective Hamiltonian for the dynamics in presence of a local magnetic field is derived. Afterwards, the magnetization of the impurity spin is investigated in Sec.~III. Sec.~IV is devoted to an analysis of the spin-spin correlation function and the results for the dynamical spin susceptibility are presented in Sec.~V.

\section{Effective Hamiltonian}

The dynamics of a spin-1/2 coupled to a sea of electrons via an exchange interaction is described by the Kondo model:
\beq
	H=\sum_{k \eta} k \mbox{:} c_{k\eta}^\dag c_{k\eta} \mbox{:} + \sum_{i} \frac{J_i}{2} \sum_{\eta,\eta'} \, \mbox{:} \Psi_{\eta}^\dag(0) \sigma_i^{\eta,\eta'} S_i \Psi_{\eta'}(0)  \mbox{:}.
\label{AIKM}
\eeq
The operator $c_{k\eta}^\dag$ creates an electron with wave vector $k$ and spin $\eta=\ua,\da$ in the reservoir. Here, we allow for an anisotropy in the exchange interaction resulting in different couplings in z-direction, $J_z=J_\parallel$, and in the $xy$ plane, $J_x=J_y=J_\perp$. In the following, the couplings are all given in units of the noninteracting density of states. The colons $:\ldots:$ denote normal ordering with respect to the Fermi sea. The local spin operator $\vec{S}$ with components $S_i,i=x,y,z,$ is coupled to the local spin density of the conduction band electrons whose components are determined by the Pauli matrices $\sigma_i$. The electron's dispersion relation has been linearized around the Fermi level and energies are measured in units of $v_F$ relative to the Fermi energy, i.e. $v_F=1$ and $\varepsilon_F=0$. As the local scatterer is assumed to be pointlike, only s-wave scattering occurs reducing the problem to a one-dimensional one.~\cite{Leggett} For an additional magnetic field $h_*$ applied to the local spin, the Hamiltonian in Eq.~(\ref{AIKM}) acquires an extra contribution $-h_* S_z$.

In the following, we will work in the Toulouse limit of the Kondo model where the parallel coupling takes a special value $J_\parallel=2-\sqrt{2}$. The relevance of the Toulouse point in the single channel Kondo model is twofold. First, it allows for exact analytical results in a strongly correlated system. Secondly, the Toulouse point governs many generic and universal properties for the whole parameter regime including the experimentally relevant isotropic case. Especially, it has been shown that the local spin dynamics, that is also investigated in this work, shows the generic behavior in equilibrium as well as out of equilibrium. Other universal quantities, however, such as the Wilson ratio explicitly depend on the anisotropy.

In the following, two non-equilibrium scenarios will be investigated that turn out to induce the same dynamics. The first is an interaction quench in the Kondo model in presence of a magnetic field, i.e. the coupling $J$ in the Kondo Hamiltonian is suddenly switched on while a local magnetic field is acting on the local spin. The second scenario investigates the dynamics in the Kondo model if the system is initially prepared in a state with a macroscopic spin polarization. 

In equilibrium, the anisotropic Kondo Hamiltonian in the Toulouse limit can be mapped onto an exactly solvable resonant level model using bosonization and refermionization.~\cite{Leggett} As has been shown by Lobaskin and Kehrein, such a mapping also exists in the case of an interaction quench where an additional local potential scattering term emerges.~\cite{Lob_dyn} Below, the implementation of the bosonization technique for the Kondo Hamiltonian with an applied local magnetic field will be presented. The bosonization technique is based on the bosonization identity
\beq
	\psi_\eta(x)=\frac{1}{\sqrt{a}} F_\eta e^{-i\frac{2\pi}{L}N_\eta x}e^{-i\phi_\eta(x)}, \:\:\: \eta=\ua\da,
\label{eq_bosonization_identity}
\eeq
that establishes a connection between fermionic fields $\psi_\eta(x)$ and bosonic fields $\phi_\eta(x)$ as an elementary operator identity in Fock space, see Ref.~\cite{Delft_Bos} for a recent review. Here, $\phi_\eta(x)=-\sum_{q>0}[e^{-iqx}b_{q\eta}+e^{iqx}b_{q\eta}^\dag]e^{-aq/2}/\sqrt{n_q}$, $b_{q\eta}=-i/\sqrt{n_q}\sum_k c_{k-q \eta}^\dag c_{k\eta}$, $q=2\pi n_q/L$ with $n_q\in \mathbb{N}$, $L$ is the system size and $a^{-1}>0$ is an ultraviolet cutoff. The Klein factor $F_\eta$ accounts for the annihilation of one electron as this cannot be accomplished by the bosonic field $\phi_\eta(x)$.

As usual, the bosonized Hamiltonian in the Toulouse limit can be simplified tremendously by applying a sequence of unitary transformations. First, the charge and spin (c,s) degrees of freedom are separated by introducing $\phi_{c/s}(x)=\frac{1}{\sqrt{2}}\left[ \phi_\ua(x) \pm \phi_\da(x) \right], \:\: \hat{N}_{c/s}=\frac{1}{2}\left[ \hat{N}_\ua \pm \hat{N}_\da \right]$. The charge sector decouples from the impurity problem and will be neglected in the following. 

In the Toulouse limit where $J_\parallel=2-\sqrt{2}$, the Emery-Kivelson transformation $U=e^{i[\sqrt{2}-1] \phi_s(0)[S_z-1/2]}$ eliminates the many-body interaction term in the Kondo Hamiltonian that couples to the $S_z$ operator. Refermionization of the transformed Hamiltonian reduces the problem to a quadratic and therefore exactly solvable one. For that purpose, another unitary transformation $U_2=e^{i\pi \hat{N}_s S_z}$ has to be imposed.~\cite{Zarand_Delft} This allows to define the fermionized spin operator $d=e^{-i\pi [\hat{N}_s-S_z]}S_-$ and its hermitian conjugate $d^\dag$ as well as new spinless fermionic fields
\begin{eqnarray}
	\psi(x)=\frac{1}{\sqrt{a}}\:F_s\: e^{-i\frac{2\pi}{L}\hat{N}_s x} \: e^{-i\phi_s(x)}, \: F_s=F_\da^\dag F_\ua,
\end{eqnarray}
with modes $c_k=(2\pi L)^{-1/2}\int dx\: e^{ikx}\: \psi(x)$. In the case of an applied local magnetic field $h_*$, an additional contribution $-h_* S_z$ appears in the Hamiltonian. The $S_z$ operator commutes with the unitary transformations $U$ and $U_2$ and can be related to the fermionic $d$ operators in the following way:
\beq
	S_z=d^\dag d-\frac{1}{2}.
\label{eq_Sz_dd}
\eeq
Thus, a local magnetic field induces a shift of the energy of the $d$ fermion. After the mapping, the effective Hamiltonian reduces to a noninteracting resonant level model with an additional local scatterer:
\begin{eqnarray}
	H_{\textnormal{\tiny eff}}& = & \sum_{k}k\mbox{:}c_k^\dag c_k \mbox{:} - \varepsilon_0 d^\dag d + g \sum_{kk'}\mbox{:}c_k^\dag c_{k'}\mbox{:} \nonumber \\
	& & \quad+ V \sum_k \left[ c_k^\dag d+d^\dag c_k \right] + \Delta E
\label{eq_effective_Hamiltonian}
\end{eqnarray}
where $V=J_\perp\sqrt{\pi/(2aL)}$, $g=[\sqrt{2}-1]\pi/L$, $\varepsilon_0=h_*$ and $\Delta E$ denotes a constant energy shift. The magnetic field solely enters the Hamiltonian by shifting the local $d$ level $\varepsilon_0$ away from the Fermi level.

The Kondo scale can be related to parameters of the resonant level model Hamiltonian by the impurity contribution to the Sommerfeld coefficient in the specific heat:~\cite{Lob_dyn} $C_{\textrm{\tiny imp}}=\gamma_{\textrm{\tiny imp}}T$ with $\gamma_{\textrm{\tiny imp}}=w\pi^2/3T_K$. Here, $w=0.4128$ is the Wilson number. In this way the Kondo temperature is determined by $T_K=\pi w \Delta$ where $\Delta=V^2 L/2$.

As will be shown in the following, a Hamiltonian of the same structure as in Eq.~(\ref{eq_effective_Hamiltonian}) is induced by preparing the system in a state with a macroscopic spin polarization:
\beq
	|\psi_0\rangle=\prod_{0<k<k_*} c_{k\ua}^\dag |0\rangle \otimes |\chi\rangle .
\label{eq_initial_state}
\eeq
Here, $|0\rangle$ is the filled Fermi sea and $|\chi\rangle$ a wave function for the local spin. For simplicity, we set the initial spin wave function $|\chi \rangle=|\ua \rangle$ in the following. The wave vector $k_*$ denotes the electronic state up to which the spin up electrons are filled. Indeed, the state $|\psi_0\rangle$ carries a macroscopic spin polarization:
\beq
	\hat{N}_s|\psi_0\rangle=\frac{L}{4\pi}k_*|\psi_0\rangle.
\label{eq_spin_polarization}
\eeq
The operator $\hat{N}_s=[\hat{N}_\ua-\hat{N}_\da]/2$ measures the total spin of the conduction band electrons.

Note, that Kondo impurities attached to ferromagnetic leads~\cite{Ferromagnetic} describe a different setup, as there the chemical potentials for the up and down spin electron species are identical, whereas the corresponding densities of states are different. Here, the chemical potentials differ in the initial state and it is assumed that the densities of states are identical by choosing the same Fermi velocities for the up and down spin electrons. This is reasonable as long as the wave vector $k_*$ is small enough in order to neglect the influence of curvature on the dispersion relation. 

Using the bosonization technique for the mapping onto a quadratic effective Hamiltonian, it is important to recognize that expressions like
\beq
	\lim_{L\to \infty} \frac{2\pi}{L} \hat{N}_s \not= 0
\label{eq_therm_limit}
\eeq
do not vanish in the thermodynamic limit due to the macroscopic initial spin polarization. Those terms usually appear as finite-size corrections when expressing the fermionic density  in terms of the bosonic fields and the number operators and cannot be neglected in the present study. As has been shown recently, the application of the bosonization technique requires additional effort in non-equilibrium scenarios.~\cite{Gutman,Millis} The implications of \Eq{eq_therm_limit} will be studied in the following for the kinetic energy of the spin sector $H_{0s}$ that equals in terms of the bosonic field $\phi_s$ and the number operator $\hat{N}_s$:
\beq
	H_{0s}=\int dx \: \frac{1}{2} :\left( \phi_s(x) \right)^2: + \frac{2\pi}{L} \hat{N}_s^2.
\label{H0s}
\eeq
Due to the spin-conservation condition~\cite{Zarand_Delft}
\beq
	S_T=\hat{N}_s+S_z=\textrm{const.}=\frac{L}{4\pi}k_*+\frac{1}{2} \stackrel{L\to\infty}{\longrightarrow}\frac{L}{4\pi} k_*
\label{spin_conservation}
\eeq
one can express the latter term of the right-hand side in \Eq{H0s} in terms of the total spin $S_T$ and the $S_z$ operator such that:
\beq
	\frac{2\pi}{L} \hat{N}_s^2=\frac{2\pi}{L}S_T^2-k_* S_z+\frac{\pi}{2L}=-k_* S_z +\textrm{const.} 
\label{eq_magnfield_kinenergy}
\eeq
Here, the identity $S_z^2=1/4$ has been used. Only in the case of a macroscopic spin-polarized initial state, one therefore obtains an additional non-vanishing contribution that is equivalent to a magnetic field of strength $k_*$ applied to the local spin. As in this example, the operator $\hat{N}_s$ can always be converted into a constant and the local spin operator $S_z$ generating new terms in the Hamiltonian compared to the equilibrium or interaction quench case without magnetic field. In the end one arrives at an effective Hamiltonian that is equal to Eq.~(\ref{eq_effective_Hamiltonian}) with $\varepsilon_0=\frac{2\pi}{L}[1-J_\parallel] S_T=[\sqrt{2}-1]k_*/2$

As shown in \ref{App_A}, the occupation distribution $f_k=\langle \psi_0 | c_k^\dag c_k | \psi_0 \rangle$ of the spinless fermionic operators in the  initial state equals:
\beq
	f_k=\theta\left( \frac{k_*}{2} - k \right).
\eeq
The chemical potential $k_*/2$ in the initial state can be compensated by defining new operators $a_k=c_{k-k_*/2}$ in terms of which the structure of the effective Hamiltonian does not change. All $c_k$ operators are replaced by $a_k$'s, only the energy of the local level gets modified to $\tilde{\varepsilon}_0=k_*/\sqrt{2}$. Therefore, the initial state effectively acts as a magnetic field of strength:
\beq
	h_*=\frac{k_*}{\sqrt{2}}.
\eeq
Notice that the magnetic field seen by the impurity spin depends on $J_\parallel$, that characterizes the strength of the coupling between the z component of the local spin and the z component of the spin-polarization cloud generated by the itinerant electrons. In the Toulouse limit, the parallel coupling is fixed to a comparatively large value $J_\parallel=2-\sqrt{2}$. We expect, however, that independent of the actual coupling strength and away from the Toulouse limit the spin-polarized initial state always acts as a local magnetic field. Moreover, even in the limit $J_\parallel \to 0$, there is an effective magnetic field stemming from the kinetic energy of the spin sector, see. Eq.~(\ref{eq_magnfield_kinenergy}). The only relevant quantity is the total effective magnetic field $h_*$ seen by the impurity spin.

Concluding, it has been shown that the dynamics in the Kondo model for an initially spin-polarized state of the type in Eq.~(\ref{eq_initial_state}) is equivalent to an interaction quench in presence of a magnetic field, see Eq.~(\ref{eq_effective_Hamiltonian}).

\section{Magnetization}

The local spin dynamics for an interaction quench in the Kondo model without magnetic field is well studied in the literature. In such a scenario, the magnetization $P_{h_*=0}(t)=\langle S_z(t) \rangle_{h_*=0}$ of the impurity spin~\cite{Lob_dyn,Leggett,Lesage,Guinea}
\beq
	P_{h_*=0}(t)=P(0)\: e^{-2\Delta t}
\label{eq_magnetization_quench}
\eeq
decays to zero exponentially fast on a time scale $1/2\Delta$ that is set by the Kondo time scale $t_K=1/T_K=1/(\pi w \Delta)$.

\begin{figure}
\centering
	\includegraphics[width=0.6\columnwidth]{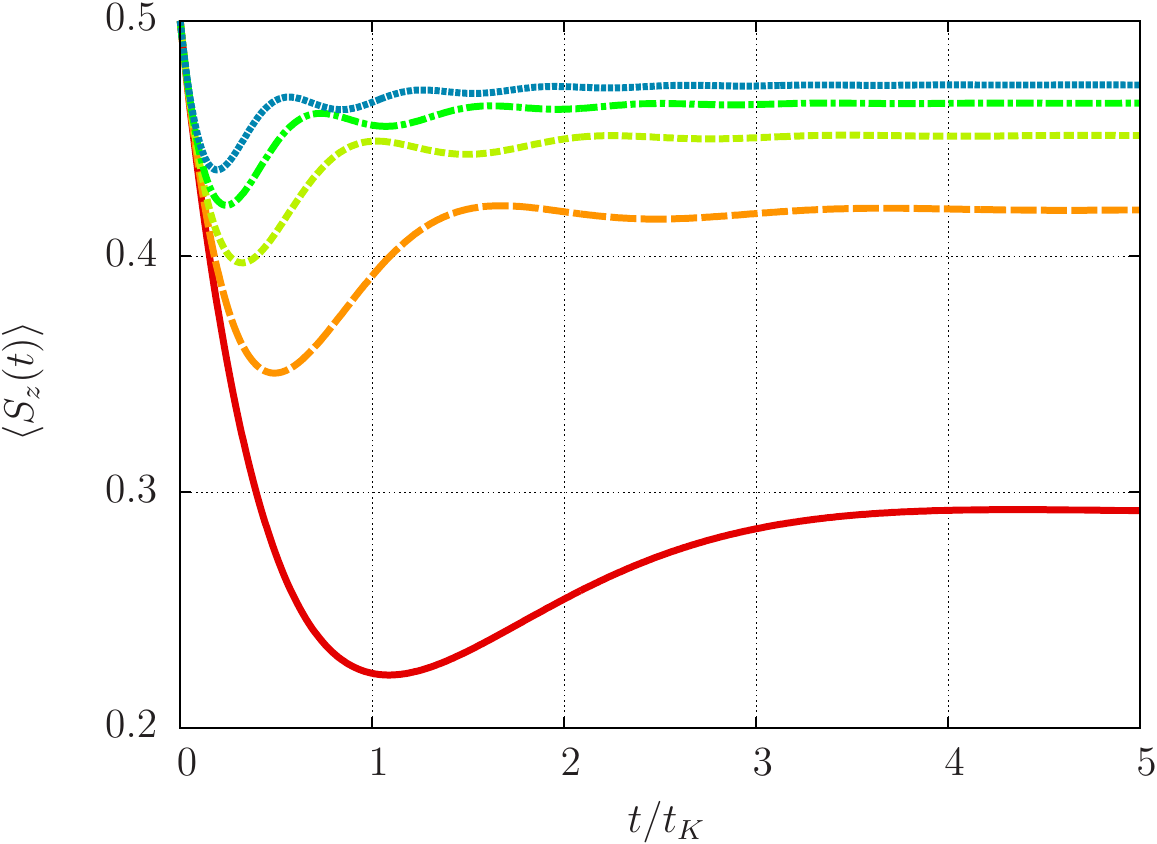}
\caption{Magnetization $P(t)=\langle S_z(t) \rangle$ for different values of the local magnetic field $h_*/T_K=9,7,5,3,1$ from top to bottom.}
\label{pic_magnetization}
\end{figure}

As the local spin is sensitive to a local magnetic field, one expects that the magnetization dynamics differs considerably compared to the interaction quench case. Due to \Eq{eq_Sz_dd}, the magnetization $P(t)$ is equivalent to the local $d$ level occupation $\hat{n}_d=d^\dag d$ up to a constant:
\beq
	P(t)=\langle \hat{n}_d(t) \rangle-\frac{1}{2}.
\eeq
As the effective Hamiltonian in \Eq{eq_effective_Hamiltonian} is quadratic, the time evolution of the single-particle operators $c_k$ and $d$ is entirely determined by the Green's functions $G_{ll'}(t)=\theta(t)\langle \{ c_{l}(t),c_{l'} \}$:
\beq
	c_l(t)=\sum_{l'} G_{ll'}(t) c_{l'}, \: l,l' =k,d
\eeq
where $G$ is a unitary matrix. The Green's functions $G_{ll'}(t)$ can be obtained by using the equations of motion approach, for example. As a result, one obtains for the magnetization of the impurity spin:
\beq
	P(t)=\frac{1}{2}\left[1+\Lambda(0) \right] e^{-2\Delta t}-e^{-\Delta t}\Lambda(t) + \frac{1}{2} \Lambda(0)
\label{eq_magnetization}
\eeq
where $\Lambda(t)=(2/\pi) \int_0^{h_*/\Delta} d\omega\: \cos(\omega \Delta t)/(1+\omega^2)$ and $\Lambda(0)=(2/\pi) \: \textrm{atan}(h_*/\Delta)$. Notice that this agrees with the result in Eq.~(\ref{eq_magnetization_quench}) in the limit $h_*\to 0$ as expected. Plots of the magnetization for different values of the magnetic field $h_*$ are shown in Fig.~\ref{pic_magnetization}. As one can see, the time scale for relaxation of the magnetization is exclusively set by the Kondo scale $t_K$ independent of the magnetic field. For the local level occupation in an interaction quench in a Majorana resonant level model, the equivalent relaxation behavior has been observed in a recent work by Komnik.~\cite{Komnik} The time scale $t_*=1/h_*$ associated with the magnetic field dominates the transient dynamics as long as $t_*<t_K$, i.e. the magnetic field $h_*>T_K$ is sufficiently large.

As the curves in Fig.~\ref{pic_magnetization} indicate, the initial decay of the magnetization is universal in the sense that it is independent of the magnetic field or equivalently the initial spin polarization and exactly equals the interaction quench case:
\beq
	\langle S_z(t) \rangle \stackrel{t\ll t_*,t_K}{\approx} \frac{1}{2} - \Delta t  + \mathcal{O}(t^2).
\eeq
The impurity spin is flipped without recognizing the presence of the magnetic field or the spin-polarized lead at short times. Approaching the time scale $t_*$, one observes a deviation from this universal initial decay and $P(t)$ reaches its minimum leading to overshooting. For sufficiently large magnetic fields $h_* > T_K$ and for times $t>t_*$, damped oscillations appear and the magnetization finally saturates at a value $P(t=\infty)=\pi^{-1}\textrm{atan}(h_*/\Delta)$, that is the equilibrium value of the magnetization in the Kondo model with an applied magnetic field $h_*$.

In terms of the initial spin polarization, the finite asymptotic value of the magnetization indicates that the impurity is not able to depolarize the system although it can flip spins. Due to spin conservation, compare \Eq{spin_conservation}, the subspace of the total Hilbert space $\mathcal{H}_{\textrm{\tiny dyn}}$ that is accessible by the dynamics can be restricted to:\cite{Zarand_Delft}
\beq
	\mathcal{H}_{\textrm{\tiny dyn}}=\mathcal{H}_+ \oplus \mathcal{H}_- 
\eeq
where the subspaces $\mathcal{H}_{\pm}$ 
\begin{eqnarray}
	\mathcal{H}_{+}=\textrm{span} \left[ \left|S_T+\frac{1}{2}, \Downarrow \right\rangle \right], \nonumber \\
	\mathcal{H}_{-}=\textrm{span} \left[ \left|S_T-\frac{1}{2}, \Uparrow \right\rangle \right]
\end{eqnarray}
contain those states that have a sea of conduction band electrons with spin $S_T \pm \frac{1}{2}$ and a local spin on the impurity with spin $\Downarrow, \Uparrow$. The impurity is therefore able to flip spins, the spin conservation condition, however, limits the maximally possible induced change of spin in the conduction band to one spin flip process from spin-up to spin-down such that the impurity is not able to compensate for the macroscopic initial spin polarization.

For large magnetic fields, $h_*\gg T_K$, the magnetization for times $t\gg t_*=1/h_*$
\beq
	\delta P(t)\stackrel{h_*\gg T_K}{\longrightarrow} \left[ \Lambda(0)-\frac{1}{2}\right] e^{-2\Delta t}-\frac{2\Delta}{\pi h_*} \frac{\sin(h_* t)}{h_* t}e^{-\Delta t}
\eeq
reveals the frequency $\Omega_*=h_*$ of the oscillations in the magnetization that can be clearly seen in Fig.~(\ref{pic_magnetization}). Here $\delta P(t)=P(t)-P(t=\infty)$ denotes the deviation from the asymptotic equilibrium value $P(t=\infty)$. 

For arbitrary fields $h_*\not=0$, the asymptotic long-time behavior of the magnetization
\beq
	\delta P(t)\stackrel{t\gg t_*,t_K}{\longrightarrow} -\frac{2}{\pi}\frac{\Delta ^2}{\Delta^2+h_*^2}\frac{\sin(h_*t)}{\Delta t} e^{-\Delta t}
\label{eq_magnetization_long_time}
\eeq
changes significantly compared to the interaction quench case without magnetic field, see \Eq{eq_magnetization_quench}, where it is exponential with rate $2\Delta$. Even a small magnetic field slows down the decay. It becomes exponential at a rate $\Delta$ times an algebraic contribution proportional to $(\Delta t)^{-1}$. Recently, a similar slow asymptotic decay for the magnetization in presence of a magnetic field was found in the work by Ratiani and Mitra~\cite{Mitra} where the transient dynamics in a quench from a single to a two-channel Kondo model has been studied. Regarding Eq.~(\ref{eq_magnetization_long_time}), in the limit $h_*\to 0$ the asymptotic long-time behavior of the magnetization crosses over into the interaction quench dynamics without magnetic field, compare Eq.~(\ref{eq_magnetization_quench}), since the prefactor vanishes. The decay becomes proportional to $e^{-2\Delta t}$ as one can see from Eq.~(\ref{eq_magnetization}) with $\Lambda(t)=0$ for $h_*=0$.

In equilibrium as well as for an interaction quench, it has been shown that the Toulouse limit of the Kondo model describes qualitatively correct the dynamics of the local spin observables.~\cite{Leggett,Lob_dyn,Lob_eff} One may raise the question whether this extends to the present case with an additional magnetic field.  In a recent work, Anders and Schiller~\cite{Anders} determined the magnetization for the same setup in the experimentally relevant case of an isotropic Kondo model at small couplings using the time-dependent numerical renormalization group technique. Comparing their numerically exact results with the analytical treatment shown here one finds very good qualitative agreement. As in the present Toulouse limit analysis, they observe an initial universal decay. Approaching $t\sim t_*$, the magnetization develops a minimum and damped oscillations appear for $t>t_*$ as long as the magnetic field is sufficiently large. The magnetization relaxes on the time scale $t_K$ to a finite asymptotic value. Those main features all appear in the present analysis such that one can expect that the behavior of the other local dynamical quantities such as the spin-spin correlation function and the dynamical spin susceptibility obtained on the Toulouse line are describing qualitatively the correct behavior for the isotropic Kondo model at small couplings.

\section{Spin-spin correlation function}

Additional information about the local dynamical properties of the system is contained in the spin-spin correlation function
\beq
	\langle S_z(t) S_z(t_w) \rangle = C(t,t_w)-\frac{i}{2} \chi(t,t_w)
\eeq
that probes the correlation between two spin measurements at different times. Its real part $C(t,t_w)=\frac{1}{2}\langle\{S_z(t),S_z(t_w) \} \rangle$ is connected to the strength of the spin fluctuations and its imaginary part determines the response function $\chi(t,t_w)=i\theta(t-t_w) \langle [ S_z(t),S_z(t_w) ] \rangle$ for times $t$ bigger than the waiting time $t_w$.

Due to \Eq{eq_Sz_dd}, the spin-spin correlation function can be related directly to operators of the effective Hamiltonian:
\beq
	\langle S_z(t) S_z(t_w) \rangle = \langle \hat{n}_d(t) \hat{n}_d(t_w) \rangle - \frac{1}{2 }\left[ P(t) + P(t_w) \right] -\frac{1}{4}.
\eeq
From the evaluation of $\langle \hat{n}_d(t) \hat{n}_d(t_w) \rangle$ one obtains for the cumulant $\langle S_z(t) S_z(t_w) \rangle_C=\langle S_z(t) S_z(t_w) \rangle-\langle S_z(t) \rangle \langle S_z(t_w) \rangle$ of the spin-spin correlation function:
\begin{eqnarray}
	\langle S_z(t) S_z(t_w) \rangle_C=\Omega(t,t_w) \left[ \Lambda(t-t_w)-e^{-\Delta t}\Lambda(t_w)- \right. \nonumber \\
	\left.-e^{-\Delta t_w} \Lambda(t) +e^{-\Delta(t+t_w)} (1+\Lambda(0)) +\Omega(t,t_w)\right]
\end{eqnarray}
where the function $\Omega$ is given by:
\beq
	\Omega(t,t_w)=\frac{1}{\pi} \int_{h_*/\Delta}^\infty d\omega \: \frac{\left[ e^{-i\omega \Delta t} - e^{-\Delta t} \right] \left[ e^{i\omega \Delta t_w} - e^{-\Delta t_w} \right]}{1+\omega^2}.
\eeq
One fundamental question connected to non-equilibrium quench dynamics concerns the thermalization behavior of observables.~\cite{Lob_dyn,Lob_eff,quenches,Nordl_Kondo_develop,Anders,Kondo_neq,Mitra} As for the magnetization, the relaxation of the spin-spin correlation function happens exponentially fast on a time scale $1/\Delta\propto t_K$ set by the Kondo scale with a further suppression by an additional algebraic contribution $(\Delta t_w)^{-1}$:
\beq
	\delta \langle S_z(t+t_w)S_z(t_w) \rangle \stackrel{t_w \gg t_*,t_K}{\longrightarrow} F(t,t_w)\frac{e^{-\Delta t_w}}{\Delta t_w}.
\label{eq_relax_spsp}
\eeq
Here, $\delta\langle S_z(t+t_w)S_z(t_w) \rangle=\langle S_z(t+t_w)S_z(t_w) \rangle-\langle S_z(t)S_z \rangle_{eq}^{h_*}$ denotes the deviation of the spin-spin correlation function from its asymptotic relaxed form which is just the equilibrium spin-spin correlation function $\langle S_z(t)S_z \rangle_{eq}^{h_*}$ with applied magnetic field $h_*$. The function $F(t,t_w)$ is an oscillating function of $t_w$ with period $2\pi/h_*$ and does not contribute to the relaxation dynamics. Comparing \Eq{eq_relax_spsp} with the result for an interaction quench obtained by Lobaskin and Kehrein~\cite{Lob_dyn}, one observes that the relaxation behavior of the spin-spin correlation function does not change considerably. The magnetic field only leads to a modification of the function $F(t,t_w)$. As already observed for the magnetization, the time scale for thermalization of the spin-spin correlation function is exclusively set by the Kondo scale $t_K$.

In equilibrium at zero temperature without magnetic field, the spin-spin correlation function shows a universal algebraic long-time behavior~\cite{Leggett}
\beq
	\langle S_z(t) S_z \rangle_{eq}\stackrel{t\gg t_K}{\longrightarrow} -\frac{1}{[\pi \Delta t]^2}.
\label{eq_asympt_eq}
\eeq
In the present non-equilibrium setting, the universal long-time behavior remains unchanged:
\begin{eqnarray}
	\langle S_z(t) S_z(t_w) \rangle_{\textrm{C}}\stackrel{t\gg t_K,t_*,t_w}{\longrightarrow} -\left[ \frac{\Delta^2}{h_*^2+\Delta^2} \right]^2 \times \nonumber \\
	 \times \frac{1-2\cos(h_* t_w)e^{-\Delta t_w}+e^{-2\Delta t_w}}{[\pi \Delta (t-t_w)]^2}
\label{eq_spsp_longtime}
\end{eqnarray}
with only a different prefactor. This universality originates from the property of the spin-spin correlation function that its asymptotic long-time behavior is solely dependent on the low-energy excitations in the vicinity of the Fermi level as one can show by a simple Fermi liquid argument. In equilibrium, one can expand the $d$ operator in the energy representation of operators $a_{\varepsilon}$ that diagonalize the Hamiltonian in Eq.~(\ref{eq_effective_Hamiltonian}), such that one obtains:
\beq
	d=\rho_0 ^{-1/2}\sum_{\varepsilon} f_\varepsilon a_\varepsilon .
\eeq
Here, $|f_\varepsilon|^2$ is the local single-particle density of states $\rho_d(\varepsilon)$ of the effective Hamiltonian at energy $\varepsilon$ and $\rho_0=L/(2\pi)$ is the noninteracting density of states of the lead. In the energy representation, the spin-spin correlation function then reduces to:
\beq
	\langle S_z(t) S_z \rangle_{eq}=\left[ \rho_0^{-1} \sum_{\varepsilon>0} |f_\varepsilon|^2 e^{-i \varepsilon t} \right]^2.
\eeq
Performing a Wick rotation $t\to-i\tau$, one observes that for large $\tau$ only the low energy states contribute. Assuming that $|f_\varepsilon|^2$ is well behaved near the Fermi level $\varepsilon=0$ such that it may be expanded around $\varepsilon=0$ and replacing $\tau$ by $i t$, the long-time behavior of the spin-spin correlation function is given by:
\beq
	\langle S_z(t) S_z \rangle_{eq} \stackrel{t \to \infty}{\longrightarrow} \left[ \frac{\rho_d}{ i t} \right]^2.
\eeq
In equilibrium at zero magnetic field, the local level $\varepsilon_0=0$ exactly lies at the Fermi energy such that the local density of states $\rho_d$ at the Fermi level equals $\rho_d=1/(\pi \Delta)$ and the exact result of Eq.~(\ref{eq_asympt_eq}) is reproduced. In case of an applied magnetic field $h_*$, the local $d$ level is shifted away from the Fermi energy leading to a reduction of the local density of states $\rho_d=(\Delta/\pi)/(\Delta^2+h_*^2)$. Comparing with the result in Eq.~(\ref{eq_spsp_longtime}), one observes that the prefactor in the long-time behavior of the relaxed spin-spin correlation function for $t_w\to\infty$ exactly equals the square of the local density of states as expected from the Fermi liquid argument above.

For intermediate waiting times $0<t_w<\infty$, the long-time behavior acquires a modification by damped oscillations with period $2\pi/h_*$ as a function of the waiting time $t_w$. This is a consequence of the transient non-equilibrium state where the Fermi liquid argument is not valid. At zero waiting time $t_w=0$, the decay becomes exponential as the initial state is an eigenstate of the $S_z$ operator and the spin-spin correlation function reduces to the magnetization of Sec. III as has been observed for the case of an interaction quench without magnetic field.~\cite{Lob_dyn}

\section{Dynamical spin susceptibility}

\begin{figure}
\centering
\begin{multicols}{2}
\includegraphics[width=\linewidth]{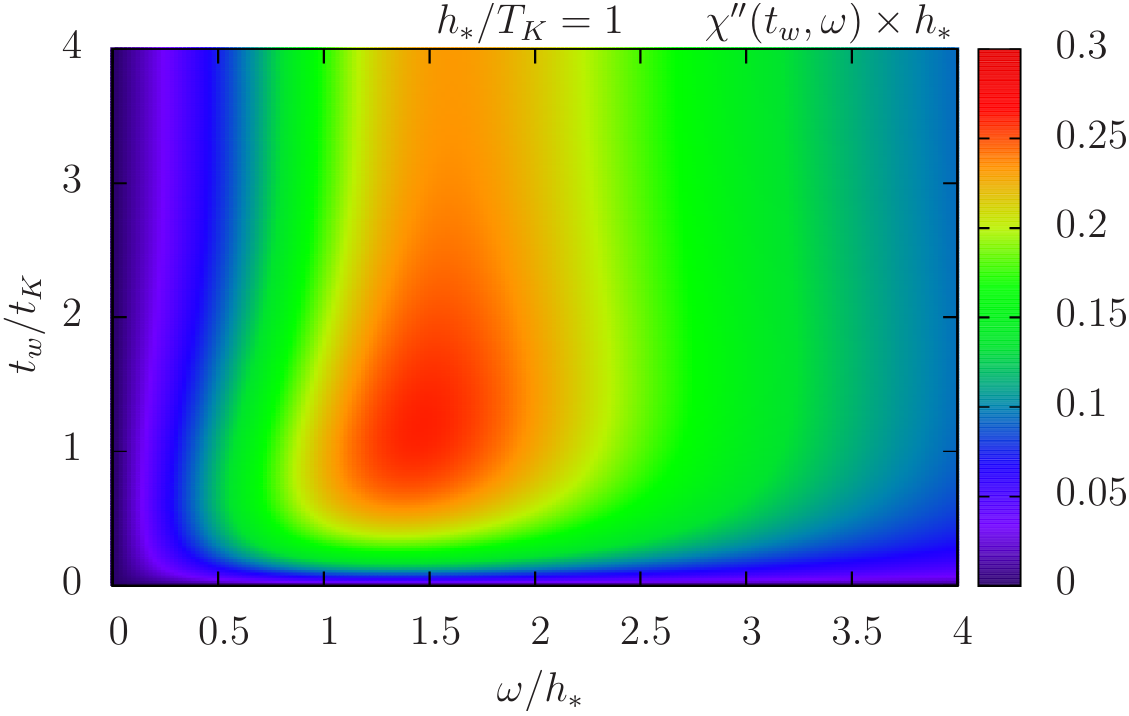}

\includegraphics[width=\linewidth]{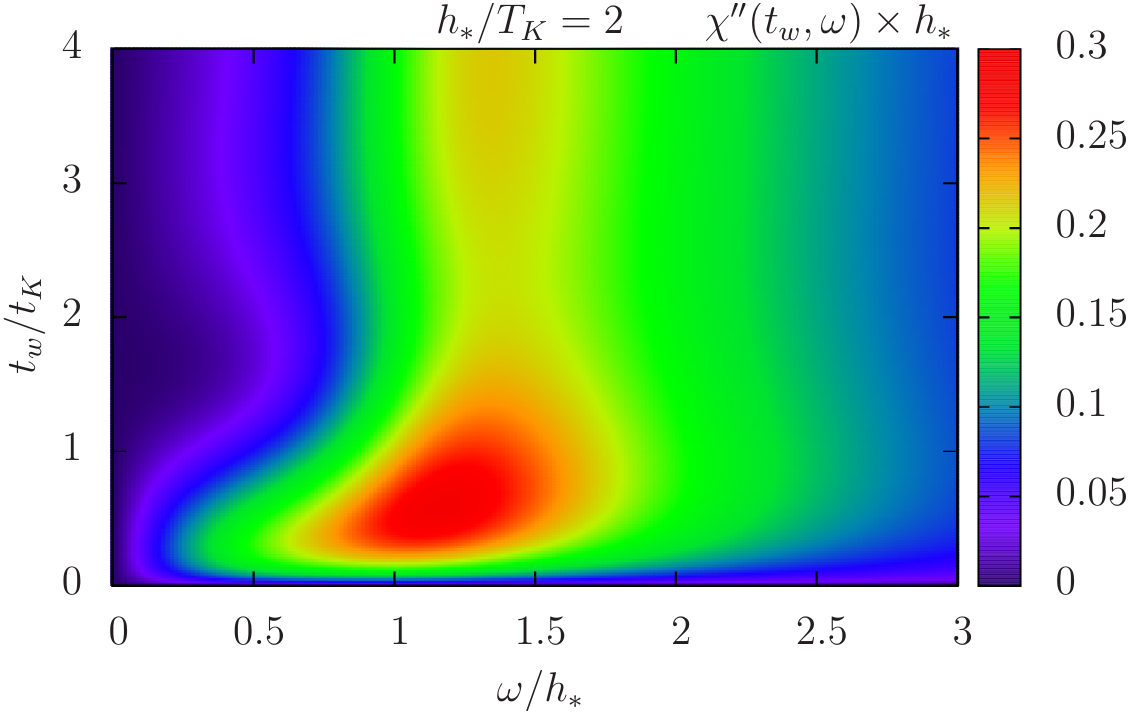}
\end{multicols}

\begin{multicols}{2}
\includegraphics[width=\linewidth]{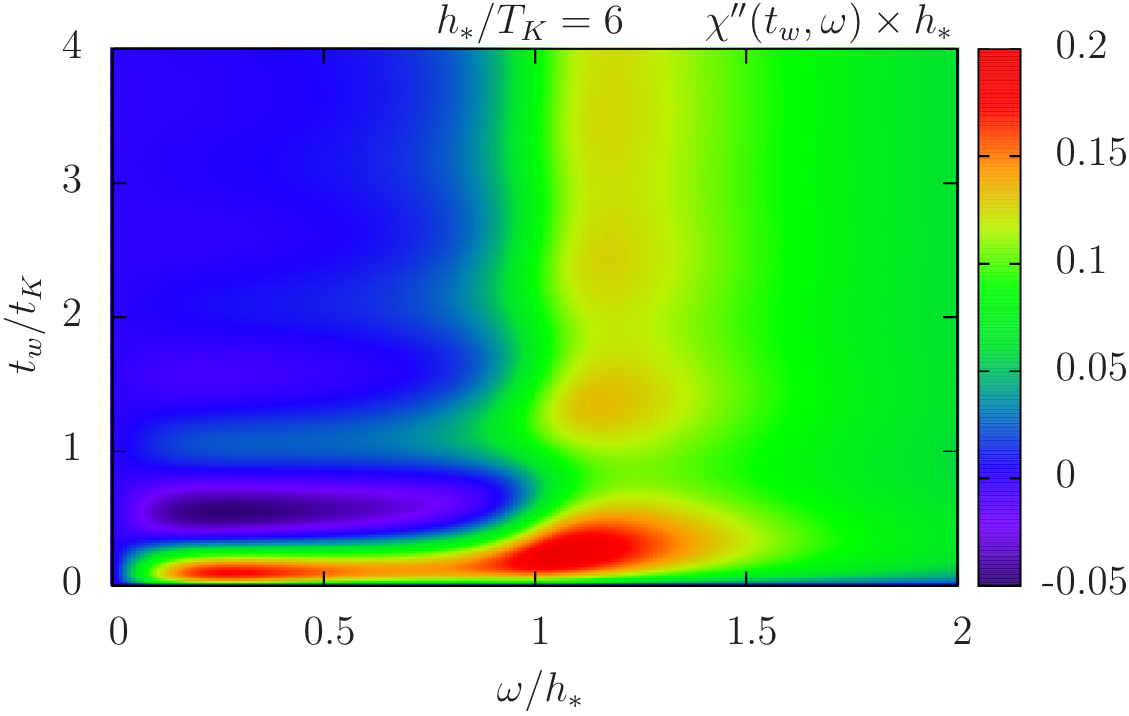}

\includegraphics[width=\linewidth]{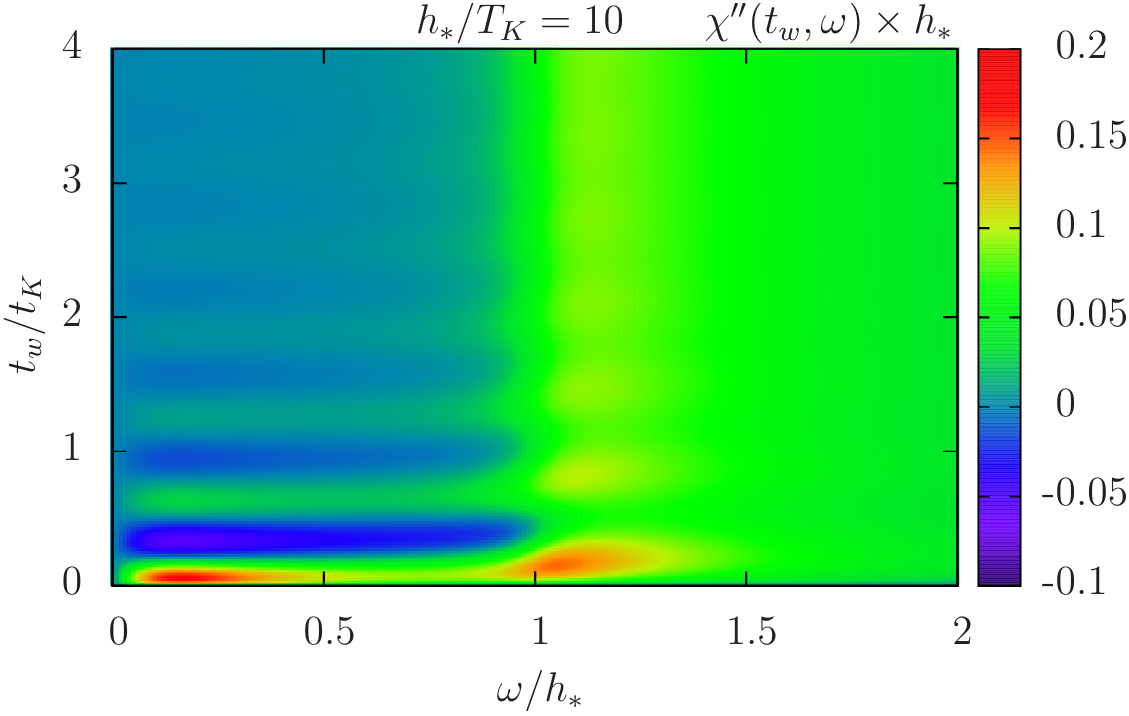}
\end{multicols}
\caption{False color plots of the dynamical spin susceptibility $\chi''(t_w,\omega)$ for different values of the magnetic field $h_*$.}
\label{pic_dynsusz}
\end{figure}
The imaginary part of the spin-spin correlation function determines the response of the system to a small external magnetic field $h(t)$ applied to the local spin. In the linear response regime, the magnetization in presence of $h(t)$ equals:
\beq
	\langle S_z(t) \rangle_h=\langle S_z(t) \rangle + \int_{-\infty}^\infty dt' \: \chi(t,t') h(t').
\eeq
Expectation values without an index $h$ are to be evaluated with the unperturbed Hamiltonian. In equilibrium, the response function depends only on the time difference thereby establishing a spectral representation in terms of one frequency. Its imaginary part $\chi''(\omega)$, the dynamical spin susceptibility, shows a peak near the Kondo temperature representing the Kondo singlet. In the non-equilibrium setting considered in this work, the suitable generalization of the spectral representation of the response function is the following:~\cite{Lob_dyn}
\beq
	\chi(t_w,\omega)=2\int_{0}^\infty \frac{dt}{2\pi} \: e^{i \omega t} \chi(t+t_w,t_w).
\eeq
The integration is only taken over the positive real axis as a consequence of causality such that an additional prefactor of $2$ is required in order to reproduce the equilibrium case. The function $\chi(t_w,\omega)$ may be interpreted as the spectral decomposition of the response function at a given point $t_w$ in time. Accordingly, $\chi''(t_w,\omega)=\textrm{Im} \chi(t_w,\omega)$ determines the dynamical spin susceptibility at a given point $t_w$ in this non-equilibrium setup. False color plots of $\chi''(t_w,\omega)$ are shown in Fig.~\ref{pic_dynsusz} for different degrees of initial spin polarization.

As one can see from Fig.~\ref{pic_dynsusz}, the time scale for relaxation of the dynamical spin susceptibility is solely determined by the Kondo scale $t_K$ as already found for the magnetization and the spin-spin correlation function.

For small magnetic fields $h_* < T_K$, the behavior of the dynamical spin susceptibility resembles the interaction quench case without magnetic field. In the transient regime up to times $t\sim t_K$, the dynamical spin susceptibility approaches its equilibrium shape exponentially fast with a peak located at $\omega \approx T_K$ associated with the Kondo singlet as one observes in the case without magnetic field.~\cite{Nordl_Kondo_develop,Lob_dyn}

For magnetic fields $h_*\gtrsim T_K$, as can be seen in the plots of Fig.~\ref{pic_dynsusz}, a maximum builds up on a time scale $t_*$ whose position roughly scales proportional to the magnetic field. Importantly, the peak associated with the Kondo singlet near $T_K$ is not visible any more indicating that the large magnetic field avoids the buildup of Kondo correlations. The spin fluctuations that are the basis for the emergence of the Kondo effect are suppressed as the impurity spin is pinned by the strong magnetic field. After a fast initial buildup on a time scale $t_*$ associated with the initial spin polarization, damped oscillations appear in the dynamical spin susceptibility with frequency $h_*$ decaying on a scale $t_K$. 

\section{Conclusions}

In this work, the quench dynamics of the Kondo model in the Toulouse limit has been analyzed in presence of a local magnetic field. It has been shown that this setup can be realized either by applying the magnetic field directly or by preparing the system in a macroscopically spin-polarized state. The effective magnetic field caused by the initial state results from a subtlety in applying the bosonization technique. Terms that are usually referred to as finite-size corrections become important in this non-equilibrium setting. The transient dynamics has been investigated by analyzing exact analytical results for the local spin dynamics such as the magnetization, the spin-spin correlation function and the dynamical spin susceptibility.

The analysis revealed that the time scale for relaxation of all the investigated local dynamical quantities is exclusively set by the Kondo scale $t_K$ and is independent of the magnetic field $h_*$, compare Eqs.~(\ref{eq_magnetization_quench}), (\ref{eq_relax_spsp}) and Fig.~(\ref{pic_magnetization}), (\ref{pic_dynsusz}).

The magnetization shows damped oscillations with a frequency $h_*$ set by the magnetic field for $h_*>T_K$. Compared to the interaction quench case without magnetic field, the asymptotic long-time decay of the magnetization becomes slower in the present setup with a different scaling behavior, compare Eq.~(\ref{eq_magnetization_long_time}). Overall, the analysis of the magnetization on the Toulouse line shows very good qualitative agreement with a recent numerically exact analysis for the isotropic Kondo model obtained from a time-dependent numerical renormalization group study.~\cite{Anders} Therefore, one can expect that the other local dynamical quantities analyzed in this work such as the spin-spin correlation function and the dynamical spin susceptibility also display the main features and that the results presented in this work are qualitatively valid also away from the Toulouse limit.

The relaxation behavior of the spin-spin correlation function, see Eq.~(\ref{eq_relax_spsp}), is not altered considerably compared to the interaction quench case without magnetic field. Remarkably, the asymptotic long-time behavior of the spin-spin correlation function is universal and its scaling behavior equals the equilibrium case. This universality originates in the property that the asymptotic long-time behavior of the spin-spin correlation function is solely dependent on the low-energy excitations in the vicinity of the Fermi level. By using a simple Fermi liquid argument, it was shown that the only relevant quantity for the prefactor is the local density of states of the quadratic effective Hamiltonian at the Fermi level.

The dynamical spin susceptibility is not influenced substantially for magnetic fields $h_*< T_K$ smaller than the Kondo temperature and behaves as in an interaction quench scenario without magnetic field. For $h_*>T_K$, one observes a rapid buildup of correlations on a time scale $t_*=1/h_*$ and damped oscillations appear for times $t>t_*$ that decay on a time scale $t_K$. The dynamical spin susceptibility shows a maximum whose position roughly scales proportional to the applied magnetic field. Most importantly, the Kondo peak near $T_K$ disappears for sufficiently large $h_*$ indicating that the magnetic field seen by the impurity spin avoids the buildup of Kondo correlations by pinning the local spin thereby suppressing the local spin fluctuations.

In summary we have obtained exact results for the interaction quench dynamics of the Kondo model on the Toulouse line and studied its thermalization behavior. This is a rare case where rigorous results can be found even in non-equilibrium and provides another benchmark in the growing field of non-equilibrium quantum impurity physics.~\cite{Lob_dyn,Lob_eff,Nordl_Kondo_develop,Anders,Kondo_neq,Mitra,Lesage,Komnik}

\ack
This work was supported through SFB~TR12 of the Deutsche 
Forschungsgemeinschaft (DFG), the Center for Nanoscience (CeNS)
Munich, and the German Excellence Initiative via the
Nanosystems Initiative Munich (NIM).

\appendix
\section{}
\label{App_A}

For the evaluation of correlation functions in the macroscopically spin-polarized initial state, it is essential to determine the initial occupation distribution of the spinless fermions
\beq
	f_{kk'}=\langle \psi_0'| c_k^\dag c_{k'}|\psi_0' \rangle
\eeq
where the state $|\psi_0'\rangle$ is given by:
\begin{eqnarray}
	|\psi_0'\rangle=e^{i\pi \hat{N}_s S_z} e^{i[\sqrt{2}-1]\phi_s(0)[S_z-\frac{1}{2}]} |\psi_0\rangle =e^{iLk_*/8}|\psi_0\rangle
\end{eqnarray}
as the initial state $|\psi_0\rangle$ is an eigenstate of both $S_z$ and $N_s$. Expressing the modes $c_k$ in terms of the fields $\psi(x)$ and by using the bosonization identity for $\psi(x)$ one obtains:
\begin{eqnarray}
	f_{kk'} & &  =  \frac{1}{2\pi a L} \int \frac{dx}{2\pi} \int \frac{dx'}{2\pi} e^{-ikx} e^{ik'x'} \times \nonumber \\
	& & \times \langle \psi_0'| e^{i\phi_s(x)} e^{i\frac{2\pi}{L}\hat{N}_s x} e^{-i\frac{2\pi}{L}\hat{N}_s x'} e^{-i\phi_s(x')} |\psi_0'\rangle
\end{eqnarray}
Due to \Eq{eq_spin_polarization}, $e^{-i\frac{2\pi}{L}\hat{N}_s x}|\psi_0\rangle=e^{-ik_*x/2}|\psi_0\rangle$ such that:
\beq
	f_{kk'}=\delta_{kk'}\theta\left(\frac{k_*}{2}-k\right).
\eeq
Therefore, the initial spin polarized state induces a shift of the chemical potential of the spinless fermionic operators.

\end{document}